\documentclass[twocolumn,10pt,aps]{revtex4}
  
\usepackage[utf8]{inputenc}
\usepackage[T1]{fontenc}

\usepackage{amssymb}
\usepackage{amsmath}
\usepackage{amsfonts}
\usepackage{graphicx}
\usepackage{yfonts}
\usepackage{color}
\usepackage[normalem]{ulem}
\usepackage{amsthm}
\usepackage{bm}
\usepackage{bbm}
\usepackage{mathtools}
\usepackage{array}
\usepackage{placeins}
\usepackage{enumitem}
\usepackage{mathtools}

\def\<{\langle}
\def\>{\rangle}

\newcommand{\Tr}{\mathrm{Tr}}

\DeclareMathAlphabet\mathbfcal{OMS}{cmsy}{b}{n}

\mathchardef\mhyphen="2D 

\newtheorem{Lemma}{Lemma}

\newtheorem{Definition}{Definition}
\newtheorem{Remark}{Remark}
\newtheorem{Proposition}{Proposition}

\begin{document}


\title{All classes of informationally complete symmetric measurements in finite dimensions}

\author{Katarzyna Siudzi\'{n}ska}
\affiliation{Institute of Physics, Faculty of Physics, Astronomy and Informatics \\  Nicolaus Copernicus University in Toru\'{n}, ul. Grudzi\k{a}dzka 5, 87--100 Toru\'{n}, Poland}

\begin{abstract}
A broad class of informationally complete symmetric measurements is introduced. It can be understood as a common generalization of symmetric, informationally complete POVMs and mutually unbiased bases. Additionally, it provides a natural way to define two new families of mutually unbiased symmetric measurement operators in any finite dimension. We show a general method of their construction, together with an example of an optimal measurement. Finally, we analyze the properties of symmetric measurements and provide applications in entropic relations and entanglement detection.
\end{abstract}

\flushbottom

\maketitle

\thispagestyle{empty}

\section{Introduction}

Quantum measurements are important tools in quantum information processing for obtaining information about quantum states. In quantum information theory, measurements are represented by a collection of positive operator-valued measures (POVMs); that is, positive semidefinite operators that sum up to the identity operator. Measurements are informationally complete if they allow for a full tomography of a quantum state \cite{Prugovecki}. In particular, if the Hilbert space of a quantum system is $d$-dimensional, then a complete information about the state is obtained from the expectation values of $d^2$ operators spanning the $d^2$-dimensional space of observables.

Through the introduction of symmetries between the measurement operators, one recovers two important classes of projective measurements. A symmetric, informationally complete (SIC) POVM consists in $d^2$ elements such that every operator inner product between different pairs is the same \cite{Renes}. Mutually unbiased bases (MUBs) are (at most) $d+1$ sets of $d$ orthogonal projectors where the transition probability between any two states from different bases is constant \cite{Schwinger,Szarek}. Interestingly, SIC POVMs and MUBs have many common applications, including quantum state tomography \cite{Adamson,Scott}, quantum key distribution \cite{Renes2,Cerf}, and quantum entanglement detection \cite{Spengler,ESIC}. They are also often analyzed in the same context \cite{Wootters3,Rastegin5,Beneduci,OperationalSICs}. This indicates that the two objects are closely related.

In this paper, we show that SIC POVMs and MUBs are special cases of a more general concept: collections of mutually unbiased symmetric measurements that form an informationally complete set. Further analysis reveals that there are indeed two more classes of such measurements in every finite dimension. We provide a general method of construction as well as an example of optimal measurement with projective operators of rank 2. Possible applications are discussed in entropic uncertainty relations and quantum entanglement detection.

Our results indicate that symmetric measurements have the ability to become just as versitile in their applications in both theoretical and experimental physics as SIC POVMs, MUBs, and their non-projective generalizations. For one, symmetric measurements can be used in quantum entanglement to formulate enhanced separability criteria \cite{ShenLi} and construct indecomposable entanglement witnesses \cite{TOPICAL}. In quantum cryptography, they may be applied to design quantum key distribution protocols with improved error and noise tolerance \cite{Bouchard}. Projective informationally complete symmetric measurements may provide examples of quantum 2-designs \cite{Scott,Graydon}, which are used for optimal quantum state estimation in quantum tomography. From a mathematical perspective, it might be possible to find analogues in the finite geometry between all classes of projective symmetric measurements. This could shed some light on the existence of maximal sets of MUBs and SIC POVMs in any dimension. On the other hand, non-projective measurements find applications in post processing tasks, like error rate estimation and error correction. Their advantage comes from preserving partial information about the measured state, including correlations with other subsystems.

\section{Symmetric measurements}

Consider a collection of $N$ POVMs, where each of them is $M$-elemental. Let us call this collection an $(N,M)$-POVM and denote its elements by $E_{\alpha,k}$, with the index range $\alpha=1,\ldots,N$ and $k=1,\ldots,M$. Now, we impose the symmetry conditions on the inner products of operators within the same POVM as well as between different POVMs. This leads to the following set of relations.

\begin{Definition}
An $(N,M)$-POVM is a set of $N$ POVMs $\{E_{\alpha,k};\,k=1,\ldots,M\}$ that satisfy the symmetry conditions
\begin{equation}\label{M}
\begin{split}
\Tr (E_{\alpha,k})&=w,\\
\Tr (E_{\alpha,k}^2)&=x,\\
\Tr (E_{\alpha,k}E_{\alpha,\ell})&=y,\qquad \ell\neq k,\\
\Tr (E_{\alpha,k}E_{\beta,\ell})&=z,\qquad \beta\neq\alpha.
\end{split}
\end{equation}
\end{Definition}

In Appendix \ref{AppA}, it is shown that this definition implies that
\begin{equation}\label{xyz}
w=\frac dM,\qquad y=\frac{d-Mx}{M(M-1)},\qquad z=\frac{d}{M^2},
\end{equation}
and $x$ is a free parameter from the range
\begin{equation}\label{x}
\frac{d}{M^2}<x\leq\min\left\{\frac{d^2}{M^2},\frac{d}{M}\right\}.
\end{equation}
If $x=d^2/M^2$, then the POVMs are projective measurements. On the other hand, $x=d/M^2$ corresponds to $E_{\alpha,k}=\mathbb{I}_d/M$. Note that eq. (\ref{x}) implies that it is not always possible to obtain projective measurements from our symmetric construction.

\begin{Remark}
There are no projective $(N,M)$-POVMs for $M<d$.
\end{Remark}

The $(N,M)$-POVM is informationally complete if and only if it contains $d^2$ linearly independent elements. For every $M$-elemental POVM numbered by the index $\alpha$, there are $M-1$ linearly independent operators due to the constraint $\sum_{k=1}^ME_{\alpha,k}=\mathbb{I}_d$. Collectively, all POVMs share the final spanning operator, which is the identity $\mathbb{I}_d$. Therefore, $M$ and $N$ are related via $(M-1)N=d^2-1$. Provided that this condition holds, the set of operators $\{E_{\alpha,k}\}$ is indeed informationally complete (see Appendix \ref{AppB}). Any state $\rho$ can be decomposed into
\begin{equation}\label{rho}
\rho=\sum_{\alpha=1}^N\sum_{k=1}^Mp_{\alpha,k}F_{\alpha,k},
\end{equation}
where $p_{\alpha,k}=\Tr(E_{\alpha,k}\rho)$ are the probability outcomes and
\begin{equation}\label{Fak}
F_{\alpha,k}=\frac{1}{x-y}\left(E_{\alpha,k}-\frac{(N-1)z+y}{Nw}\mathbb{I}_d\right).
\end{equation}

Up until this point, our construction could be seen as nothing more than a common generalization of SIC POVMs and MUBs. Actually, the above considerations introduce in fact a unified method to describe general SIC POVMs \cite{Appleby,Gour} and mutually unbiased measurements (MUMs) \cite{Kalev}, which respectively generalize the notions of SIC POVMs and MUBs to nonprojective measurements. However, these provide only two special cases of $(N,M)$-POVMs. Further observations can be made based on the relation
\begin{equation}\label{mn}
(M-1)N=d^2-1,
\end{equation}
which determines the decomposition of $d^2-1$ linearly independent operators (without the identity) into $N$ POVMs with $M$ elements each. For $d=2$, there are indeed only two possible choices: $(N=1,M=4)$ for the general SIC POVM and $(N=3,M=2)$ for MUMs. However, for any finite $d>2$, there exist four decompositions of $d^2-1$ in terms of $N$ and $M$, and therefore four families of $(N,M)$-POVMs.

\begin{Proposition}\label{P1}
In any finite dimension $d$, there exist at least four informationally complete $(N,M)$-POVMs. The allowed choices of $M$ and $N$ are as follows:
\begin{enumerate}[label=(\roman*)]
\item $M=d^2$ and $N=1$ (general SIC POVM),
\item $M=d$ and $N=d+1$ (MUMs),
\item $M=2$ and $N=d^2-1$,
\item $M=d+2$ and $N=d-1$.
\end{enumerate}
\end{Proposition}

There can be, of course, more choices of $M$ and $N$, depending on the number of divisors for $d^2-1$. For example, for the first non-trivial case of $d=5$, one has two additional choices besides the standard four; namely, $(N=8,M=4)$ and $(N=3,M=9)$.

\section{Construction method}

By following the procedure introduced by Kalev and Gour \cite{Gour,Kalev}, we provide a general method of constructing any informationally complete $(N,M)$-POVM. Let us take an orthonormal Hermitian operator basis $\{G_0=\mathbb{I}_d/\sqrt{d},G_{\alpha,k};\,\alpha=1,\ldots,N,\,k=1,\ldots,M-1\}$ with traceless $G_{\alpha,k}$. Next, construct $MN$ operators
\begin{equation}\label{H}
H_{\alpha,k}=\left\{\begin{aligned}
&G_\alpha-\sqrt{M}(\sqrt{M}+1)G_{\alpha,k},\quad k=1,\ldots,M-1,\\
&(\sqrt{M}+1)G_\alpha,\qquad k=M,
\end{aligned}\right.
\end{equation}
where $G_\alpha=\sum_{k=1}^{M-1}G_{\alpha,k}$. The elements of $(N,M)$-POVMs are given by
\begin{equation}\label{E}
E_{\alpha,k}=\frac 1M \mathbb{I}_d+tH_{\alpha,k}.
\end{equation}
The parameter $t$, which is chosen in such a way that $E_{\alpha,k}\geq 0$, belongs to the range
\begin{equation}
-\frac{1}{M}\frac{1}{\lambda_{\max}}\leq t\leq \frac{1}{M}\frac{1}{|\lambda_{\min}|},
\end{equation}
where $\lambda_{\max}$ and $\lambda_{\min}$ are the minimal and maximal eigenvalue from among all eigenvalues of $H_{\alpha,k}$, respectively. This construction guarantees that eqs. (\ref{M}) and (\ref{xyz}) defining the $(N,M)$-POVM are satisfied.

The relation between $x$ and $t$ is provided by the formula
\begin{equation}\label{xt}
x=\frac{d}{M^2}+t^2(M-1)(\sqrt{M}+1)^2.
\end{equation}
Note that if $E_{\alpha,k}$ is positive for a given value of $x$, then it is also positive for any $x^\prime\in(d/M^2,x)$. Therefore, one is usually interested in determining the optimal value $x_{\rm opt}$ of $x$. Coincidentally, $x_{\rm opt}$ depends on the choice of $G_{\alpha,k}$. The closer this parameter is to $d^2/M^2$, the closer the associated POVM is to a projective measurement. If $x_{\rm opt}$ reaches the upper bound of $x$, then the corresponding $(N,M)$-POVM is called {\it optimal}. Examples include the SIC POVMs and mutually unbiased bases.

Conversely, given an $(N,M)$-POVM, one recovers the Hermitian basis
\begin{equation}
\begin{split}
G_{\alpha,k}=\frac{1}{tM(\sqrt{M}+1)^2}\Big[\mathbb{I}_d&+\sqrt{M}E_{\alpha,M}\\
&-\sqrt{M}(\sqrt{M}+1)E_{\alpha,k}\Big].
\end{split}
\end{equation}
The correspondence between $G_{\alpha,k}$ and $E_{\alpha,k}$ is one-to-one. Given any orthonormal Hermitian operator basis that consists in all traceless operators and the identity, one can always construct a corresponding $(N,M)$-POVM. Moreover, any $(N,M)$-POVM can be obtained through this method.

Observe that in general it is not easy to find the relation between two different $(N,M)$-POVMs. However, the problem simplifies significantly for a special case where $(N^\prime=M-1,M^\prime=N+1)$. This is true e.g. for the pairs {\it (i)--(iii)} and {\it (ii)--(iv)} from Proposition \ref{P1}. Then, the $(N,M)$-POVM and the $(N^\prime,M^\prime)$-POVM can be both constructed using the same orthonormal basis provided that $G_{\alpha^\prime,k^\prime}=G_{k,\alpha}$. This observation indicates that there might be a closer relation between the aforementioned families of $(N,M)$-POVMs than between SIC POVMs and MUBs.

\section{Example -- Optimal POVMs}

Let us provide one more example of an optimal informationally complete $(N,M)$-POVM; that is, a measurement for which the free parameter $x$ reaches its upper bound. Consider the special case with $N=d^2-1$ and $M=2$. Using eqs. (\ref{H}) and (\ref{E}), one finds the formula for the measurement operators
\begin{equation}
E_{\alpha,k}=\frac 12 \mathbb{I}_d +t(-1)^k(1+\sqrt{2})G_\alpha,
\end{equation}
where $G_\alpha$ are the traceless Hermitian basis operators. From eq. (\ref{xt}), we see that the upper bound $x=d/2$ corresponds to
\begin{equation}
|t|=\frac{\sqrt{d}}{2(\sqrt{2}+1)}.
\end{equation}
Moreover, $|t|=1/|2\lambda|$, where $\lambda$ is the eigenvalue of $H_{\alpha,k}=(-1)^k(1+\sqrt{2})G_\alpha$ with the greatest module. Therefore, an extremal eigenvalue of the corresponding $G_\alpha$ is $\pm 1/\sqrt{d}$. A good example of such basis is the tensor product of Pauli matrices $G_{di+j}=\sigma_i\otimes\sigma_j$ in $d=4$, where $\sigma_0=\mathbb{I}_2/\sqrt{2}$ and
\begin{equation*}
\sigma_1=\frac{1}{\sqrt{2}}\begin{pmatrix} 0 & 1 \\ 1 & 0 \end{pmatrix},\,
\sigma_2=\frac{1}{\sqrt{2}}\begin{pmatrix} 0 & -i \\ i & 0 \end{pmatrix},\,
\sigma_3=\frac{1}{\sqrt{2}}\begin{pmatrix} 1 & 0 \\ 0 & -1 \end{pmatrix}.
\end{equation*}
Then, the $(3,2)$-POVM elements, given by
\begin{equation}
E_\alpha=\frac 12 \mathbb{I}_4\pm G_\alpha,
\end{equation}
are rank-2 projectors, as $E_\alpha^2=E_\alpha$ have two non-vanishing eigenvalues.

\section{Applications}

\subsection{Entropic relations}

Although uncertainty relations were first formulated with variance, they can be alternatively expressed in terms of entropy. With the development of quantum information theory, they found new applications in quantum cryptography \cite{Koashi,Coles}, quantum entanglement \cite{Guhne2,Rastegin4}, and non-locality \cite{Oppenheim}.

To derive entropic uncertainty relations for $(N,M)$-POVMs, we first find the index of coincidence. It is defined as a sum of squared probabilities \cite{Rastegin5},
\begin{equation}
C=\sum_{\alpha=1}^N\sum_{k=1}^Mp_{\alpha,k}^2,
\end{equation}
where in our case $p_{\alpha,k}=\Tr(\rho E_{\alpha,k})$.

\begin{Lemma}\label{lemma}
Let $\{E_{\alpha,k};\,\alpha=1,\ldots,N;\,k=1,\ldots,M\}$ be an informationally complete $(N,M)$-POVM characterized by the parameter $x$. For an arbitrary state $\rho$, the index of coincidence $C$ is upper bounded by
\begin{equation}\label{coin}
\widetilde{C}=\frac{d-1}{d}\frac{d^2+M^2x}{M(M-1)}.
\end{equation}
\end{Lemma}

For a detailed proof, as well as an exact expression for $C$, refer to Appendix \ref{AppC}. Observe that for optimal projective $(N,M)$-POVMs, where $x=d^2/M^2$, the index of coincidence obeys a simpler condition,
\begin{equation}
C\leq\frac{2d(d-1)}{M(M-1)}.
\end{equation}
In this case, the greater the value of $M$, the tighter the upper bound.

Now, using Lemma \ref{lemma}, we obtain the entropic uncertainty relations (proof in Appendix \ref{AppD})
\begin{equation}
\frac{1}{N}\sum_{\alpha=1}^NH(\mathcal{E}_\alpha,\rho)
\geq\ln\frac{N}{C}
\end{equation}
for the Shannon entropy $H(\mathcal{E}_\alpha,\rho)=-\sum_{k=1}^Mp_{\alpha,k}\ln p_{\alpha,k}$ associated with the measurement $\mathcal{E}_\alpha=\{E_{\alpha,k};\,k=1,\ldots,M\}$.
The special cases agree with the entropic relations for mutually unbiased bases \cite{Ivanovic,Sanchez} and measurements \cite{ChenFei}, as well as SIC POVMs \cite{Rastegin5} and their generalization \cite{Rastegin6}.

\subsection{Entanglement detection}

Quantum entanglement is a key feature in quantum theory. It provides important resources for modern quantum technologies, such as quantum cryptography \cite{Ekert}, quantum teleportation \cite{Brassard}, and quantum computation \cite{Raus}. An important problem is to distinguish between separable and entangled states.

Consider a bipartite quantum state $\rho$ of the composite system with the underlying Hilbert space $\mathcal{H}=\mathcal{H}_A\otimes\mathcal{H}_B$, where $\dim\mathcal{H}_{A/B}=d_{A/B}$. On each subsystem, we choose an $(N_{A/B},M_{A/B})$-POVM and denote it by $\{E_{\alpha,k}^{A/B}\}$. Linear correlations between $\{E_{\alpha,k}^{A}\}$ and $\{E_{\alpha,k}^{B}\}$ are encoded in the matrix
\begin{equation}
\mathcal{P}_{\alpha,k;\beta,\ell}=\Tr\Big[\rho(E_{\alpha,k}^{A}\otimes E_{\beta,\ell}^{B})\Big].
\end{equation}
Using this matrix and Lemma 1, we construct the following separability criterion.

\begin{Proposition}\label{esic}
If a bipartite state $\rho$ is separable, then
\begin{equation}\label{es}
\|\mathcal{P}\|_{\rm tr}\leq \sqrt{\widetilde{C}_A\widetilde{C}_B},
\end{equation}
where $\widetilde{C}_{A/B}$ is the upper bound for the index of coincidence that characterizes the $(N_{A/B},M_{A/B})$-POVM.
\end{Proposition}

The proof is presented in Appendix \ref{AppE}. For SIC POVMs, one recovers the ESIC criterion \cite{ESIC}.

Now, consider a special case where $d_A=d_B=d$. Take two $(N,M)$-POVMs characterized by the same parameter $x$. Then, it is possible to formulate yet another separability criterion. Namely, if $\rho$ is a bipartite separable state, then (see Appendix \ref{AppF})
\begin{equation}
\Tr\mathcal{P}\leq\widetilde{C}.
\end{equation}
Due to $\Tr(\mathcal{P})\leq\|\mathcal{P}\|_{\rm tr}$, this criterion detects more separable states than eq. (\ref{es}). Moreover, it includes other popular separability criteria based on MUBs \cite{Spengler}, MUMs \cite{ChenMa}, and general SIC POVMs \cite{ChenLi}.

\section{Conclusions}

In this paper, we proposed the most general class of symmetric measurements that are mutually unbiased, which we call $(N,M)$-POVMs to underline the number and type of measurement operators. In particular, we found that there exist at least four informationally complete symmetric measurements in any finite dimension, including general SIC POVMs and mutually unbiased measurements. We provided a method to construct $(N,M)$-POVMs from Hermitian operator bases as well as an example of optimal measurement in dimension $d=4$. Finally, we presented possible applications in entropic uncertainty relations and entanglement detection by introducing two separability criteria.

An important open question is the existence of optimal measurements in any dimension. In other words, is it always possible to construct measurements with the maximal value of $x$. The answer is not known for MUBs nor SIC POVMs despite extensive research. A simpler task is to find optimal $(N,M)$-POVMs for the two newly introduced classes. Note that our construction can be further generalized in a manner similar to the introduction of semi-SIC POVMs \cite{Gour2}. Also, it is interesting how informationally complete symmetric measurements can be applied more efficiently than MUBs and SIC POVMs in quantum state tomography, post processing tasks, and entanglement detection.

\section{Acknowledgements}

This paper was supported by the Foundation for Polish Science (FNP) and the Polish National Science Centre project No. 2018/30/A/ST2/00837.

\bibliography{C:/Users/cynda/OneDrive/Fizyka/bibliography}

\begin{thebibliography}{10}
\providecommand{\url}[1]{\texttt{#1}}
\providecommand{\urlprefix}{URL }
\providecommand{\eprint}[2][]{\url{#2}}

\bibitem{Prugovecki}
E.~Prugove{\v{c}}ki, Int. J. Theor. Phys. \textbf{16}, 321--331 (1977).

\bibitem{Renes}
J.~M. Renes, R.~Blume-Kohout, A.~J. Scott, and C.~M. Caves, J. Math. Phys.
  \textbf{45}, 2171 (2004).

\bibitem{Schwinger}
J.~Schwinger, Proc. Nat. Acad. Sci. U.S.A. \textbf{46}, 570 (1960).

\bibitem{Szarek}
M.~B. Ruskai, S.~Szarek, and E.~Werner, Linear Algebra Appl. \textbf{347(1-3)},
  159--187 (2002).

\bibitem{Adamson}
R.~B.~A. Adamson and A.~M. Steinberg, Phys. Rev. Lett. \textbf{105}, 030406
  (2010).

\bibitem{Scott}
A.~J. Scott, J. Phys. A: Math. Gen. \textbf{39}, 13507 (2006).

\bibitem{Renes2}
J.~M. Renes, Phys. Rev. A \textbf{70}, 052314 (2004).

\bibitem{Cerf}
N.~J. Cerf, M.~Bourennane, A.~Karlsson, and N.~Gisin, Phys. Rev. Lett.
  \textbf{88}, 127902 (2002).

\bibitem{Spengler}
C.~Spengler, M.~Huber, S.~Brierley, T.~Adaktylos, and B.~C. Hiesmayr, Phys.
  Rev. A \textbf{86}, 022311 (2012).

\bibitem{ESIC}
J.~Shang, A.~Asadian, H.~Zhu, and O.~G{\"{u}}hne, Phys. Rev. A \textbf{98},
  022309 (2018).

\bibitem{Wootters3}
W.~K. Wootters, Found. Phys. \textbf{36}, 112--126 (2006).

\bibitem{Rastegin5}
A.~E. Rastegin, Eur. Phys. J. D \textbf{67}, 269 (2013).

\bibitem{Beneduci}
R.~Beneduci, T.~Bullock, P.~Busch, C.~Carmeli, T.~Heinosaari, and A.~Toigo,
  Phys. Rev. A \textbf{88}, 032312 (2013).

\bibitem{OperationalSICs}
A.~Tavakoli, M.~Farkas, D.~Rosset, J.-D. Bancal, and J.~Kaniewski, Sci. Adv.
  \textbf{7}, eabc3847 (2021).

\bibitem{ShenLi}
S.-Q. Shen, M.~Li, X.~Li-Jost, and S.-M. Fei, Quant. Inf. Proc. \textbf{17},
  111 (2018).

\bibitem{TOPICAL}
D.~Chru\'sci\'nski and G.~Sarbicki, J. Phys. A: Math. Theor. \textbf{47},
  483001 (2014).

\bibitem{Bouchard}
F.~Bouchard, K.~Heshami, D.~England, R.~Fickler, R.~W. Boyd, B.-G. Englert,
  L.~L. S{\'a}nchez-Soto, and E.~Karimi, Quantum \textbf{2}, 111 (2018).

\bibitem{Graydon}
M.~A. Graydon and D.~M. Appleby, J. Phys. A: Math. Theor. \textbf{49}, 085301
  (2016).

\bibitem{Appleby}
D.~M. Appleby, Optics and Spectroscopy \textbf{103}, 416--428 (2007).

\bibitem{Gour}
A.~Kalev and G.~Gour, J. Phys. A: Math. Theor. \textbf{47}, 335302 (2014).

\bibitem{Kalev}
A.~Kalev and G.~Gour, New J. Phys. \textbf{16}, 053038 (2014).

\bibitem{Koashi}
M.~Koashi, New J. Phys. \textbf{11}, 045018 (2009).

\bibitem{Coles}
P.~J. Coles, M.~Berta, M.~Tomamichel, and S.~Wehner, Rev. Mod. Phys.
  \textbf{89}, 015002 (2017).

\bibitem{Guhne2}
O.~G{\"u}hne, Phys. Rev. Lett. \textbf{92}, 117903 (2004).

\bibitem{Rastegin4}
A.~E. Rastegin, Quant. Inf. Proc. \textbf{15}, 2621--2638 (2016).

\bibitem{Oppenheim}
O.~G{\"u}hne, Phys. Rev. Lett. \textbf{92}, 117903 (2004).

\bibitem{Ivanovic}
I.~D. Ivanovic, J. Phys. A: Math. Gen. \textbf{25}, L363 (1992).

\bibitem{Sanchez}
J.~S{\'{a}}nchez, Phys. Lett. A \textbf{173}, 233--239 (1993).

\bibitem{ChenFei}
B.~Chen and S.-M. Fei, Quant. Inf. Proc. \textbf{14}, 2227--2238 (2015).

\bibitem{Rastegin6}
A.~E. Rastegin, Phys. Scr. \textbf{89}, 085101 (2014).

\bibitem{Ekert}
A.~K. Ekert, Phys. Rev. Lett. \textbf{67}, 661 (1991).

\bibitem{Brassard}
C.~H. Bennett, G.~Brassard, C.~Cr{\'e}peau, R.~Jozsa, A.~Peres, and W.~K.
  Wootters, Phys. Rev. Lett. \textbf{70}, 1895 (1993).

\bibitem{Raus}
R.~Raussendorf and H.~J. Briegel, Phys. Rev. Lett. \textbf{86}, 5188 (2001).

\bibitem{ChenMa}
B.~Chen, T.~Ma, and S.-M. Fei, Phys. Rev. A \textbf{89}, 064302 (2014).

\bibitem{ChenLi}
B.~Chen, T.~Li, and S.-M. Fei, Quant. Inf. Proc. \textbf{14}, 2281--2290
  (2015).

\bibitem{Gour2}
I.~J. Geng, K.~Golubeva, and G.~Gour, Phys. Rev. Lett. \textbf{126}, 100401
  (2021).

\bibitem{Maassen}
H.~Maassen and J.~B.~M. Uffink, Phys. Rev. Lett. \textbf{60}, 1103 (1988).

\end{thebibliography}
\bibliographystyle{C:/Users/cynda/OneDrive/Fizyka/beztytulow2}

\appendix

\section{Parameter values for symmetric measurements}\label{AppA}

All elements of a POVM sum up to identity. We use this to show that the values of $w$ and $z$ follow directly from the definition, as
\begin{equation}
d=\Tr(\mathbb{I}_d)=\sum_{k=1}^M\Tr(E_{\alpha,k})=Mw
\end{equation}
and
\begin{equation}
w=\Tr(E_{\alpha,k})=\sum_{\ell=1}^M\Tr(E_{\alpha,k}E_{\beta,\ell})=Mz.
\end{equation}
To obtain the formula for $y$, we compute
\begin{equation}
w=\Tr(E_{\alpha,k})=\sum_{\ell=1}^M\Tr(E_{\alpha,k}E_{\alpha,\ell})=
x+(M-1)y.
\end{equation}

The range of $x$ is determined in two steps. First, we note that the measurement operators range from rescaled rank-1 projectors to rescaled identity operators. Hence, the upper bound is $x=w^2$ and the lower bound is $x=w^2/d$. Alternatively, the lower bound follows from the Cauchy-Schwarz inequality $\Tr (E_{\alpha,k})=(\Tr E_{\alpha,k} \mathbb{I}_d) <\sqrt{dx}$, where the sharp inequality discards the case with all $E_{\alpha,k}=\mathbb{I}_d/M$.

Next, one must have $y\geq 0$ because otherwise $E_{\alpha,k}\ngeq 0$. This imposes an additional condition on $x$,
\begin{equation}
x\leq\frac{d}{M},
\end{equation}
which is stronger than $x\leq w^2=d^2/M^2$ whenever $M<d$.

\section{Proof of informational completeness}\label{AppB}

Let us take $d^2-1$ real numbers $r_{\alpha,k}$ such that
\begin{equation}\label{cond}
\sum_{\alpha=1}^N\sum_{k=1}^{M-1}r_{\alpha,k}E_{\alpha,k}=0.
\end{equation}
The $(N,M)$-POVM is informationally complete if its operators $E_{\alpha,k}$ with $\alpha=1,\ldots,N$ and $k=1,\ldots,M-1$ form a linearly independent set; that is, if eq. (\ref{cond}) implies that $r_{\alpha,k}=0$. By calculating the trace of eq. (\ref{cond}), we find that
\begin{equation}\label{cond2}
\Tr\left(\sum_{\alpha=1}^N\sum_{k=1}^{M-1}r_{\alpha,k}E_{\alpha,k}\right)
=w\sum_{\alpha=1}^N\sum_{k=1}^{M-1}r_{\alpha,k}=0.
\end{equation}
Next, we multiply eq. (\ref{cond}) by $E_{\beta,\ell}$ with an arbitrary $\beta$ before taking the trace. Using eq. (\ref{cond2}), one obtains
\begin{equation}\label{2}
\begin{split}
\Tr\Bigg(\sum_{\alpha=1}^N\sum_{k=1}^{M-1}&r_{\alpha,k}E_{\alpha,k}E_{\beta,\ell}\Bigg)
\\&=(x-y)r_{\beta,\ell}+(y-z)\sum_{k=1}^{M-1}r_{\beta,k}=0
\end{split}
\end{equation}
for $\ell=1,\ldots,M-1$, as well as
\begin{equation}\label{3}
\begin{split}
\Tr\Bigg(\sum_{\alpha=1}^N\sum_{k=1}^{M-1}r_{\alpha,k}E_{\alpha,k}&E_{\beta,M}\Bigg)
\\&=(y-z)\sum_{k=1}^{M-1}r_{\beta,k}=0.
\end{split}
\end{equation}
Taking $z=y$ results in $x=d/M^2$, which lies outside the domain of $x$. Therefore, the solution of eq. (\ref{3}) is
\begin{equation}
\sum_{k=1}^{M-1}r_{\beta,k}=0.
\end{equation}
If we input this result into eq. (\ref{2}), it follows that
\begin{equation}
(x-y)r_{\beta,\ell}=0,\qquad l=1,\dots,M-1.
\end{equation}
Coincidentally, $y=x$ also gives $x=d/M^2$ that is not an admissible value. At last, the only solution of eq. (\ref{cond}) is $r_{\beta,\ell}=0$, and hence the set $\{E_{\alpha,k}\,|\,\alpha=1,\ldots,N;k=1,\ldots,M-1\}$ indeed consists in linearly independent operators.

\section{Index of coincidence}\label{AppC}

Recall that any quantum state can be expanded into
\begin{equation}
\rho=\sum_{\alpha=1}^N\sum_{k=1}^Mp_{\alpha,k}F_{\alpha,k}
\end{equation}
using the probability distributions $p_{\alpha,k}=\Tr(\rho E_{\alpha,k})$ and the operators $F_{\alpha,k}$ given in eq. (\ref{Fak}). Now, let us calculate
\begin{equation}\label{rho2}
\Tr(\rho^2)=\sum_{\alpha,\beta=1}^N\sum_{k,\ell=1}^Mp_{\alpha,k}p_{\beta,\ell}
\Tr(F_{\alpha,k}F_{\beta,\ell}).
\end{equation}
From definition, it follows that
\begin{equation}
\Tr(F_{\alpha,k}F_{\beta,\ell})=\frac{1}{(x-y)^2}\Big[\Tr(E_{\alpha,k}E_{\beta,\ell})
+A(dA-2w)\Big],
\end{equation}
where for simplicity we introduced a new symbol
\begin{equation}
A=\frac{(N-1)z+y}{Nw}.
\end{equation}
This allows us to rewrite eq. (\ref{rho2}) into
\begin{equation*}\label{rho3}
\begin{split}
\Tr(\rho^2)&=\frac{1}{(x-y)^2}\Bigg[\sum_{\alpha=1}^N\sum_{k=1}^Mp_{\alpha,k}^2\Tr(E_{\alpha,k}^2)\\&\qquad\qquad\quad
+\sum_{\alpha=1}^N\sum_{\substack{k,\ell=1 \\ \ell\neq k}}^M
p_{\alpha,k}p_{\alpha,\ell}\Tr(E_{\alpha,k}E_{\alpha,\ell})\\&\qquad\qquad\quad
+\sum_{\substack{\alpha,\beta=1 \\ \beta\neq\alpha}}^N \sum_{k,\ell=1}^Mp_{\alpha,k}p_{\beta,\ell}\Tr(E_{\alpha,k}E_{\beta,\ell})\\&\qquad\qquad\qquad\qquad\qquad\qquad
+N^2A(dA-2w)\Bigg]\\
&=\frac{1}{(x-y)^2}\Big[Cx+(N-C)y+N(N-1)z\\&\qquad\qquad\qquad\qquad\qquad\qquad
+N^2A(dA-2w)\Big]
\end{split}
\end{equation*}
after using the properties of $E_{\alpha,k}$ from eq. (\ref{M}), as well as the definition of the index of coincidence $C$. The above formula can be further simplified via relations (\ref{xyz}) and (\ref{mn}), so that
\begin{equation*}
\begin{split}
\Tr(\rho^2)&=\frac{1}{(x-y)^2}\Big[(x-y)C+N[y+(N-1)z]\\
&\qquad\qquad\qquad\qquad\qquad+N^2A(dA-2w)\Big]\\
&=\frac{dM(M-1)C-d^3+M^2x}{d(M^2x-d)}.
\end{split}
\end{equation*}
It is straightforward to solve this equation in terms of $C$. One gets
\begin{equation}
C=\frac{d(M^2x-d)\Tr(\rho^2)+d^3-M^2x}{dM(M-1)},
\end{equation}
which is the exact (state-dependent) formula for the index of coincidence. The upper bound follows from $\Tr(\rho^2)\leq 1$.

\section{Entropic uncertainty relations}\label{AppD}

Our goal is to compute the Shannon entropy
\begin{equation}
H(\mathcal{E}_\alpha,\rho)=-\sum_{k=1}^Mp_{\alpha,k}\ln p_{\alpha,k}
\end{equation}
averaged over all measurements $\mathcal{E}_\alpha$. It has been shown that for any probability distribution \cite{Sanchez,Maassen}
\begin{equation}
\sum_{k=1}^Mp_{\alpha,k}\ln p_{\alpha,k}\leq\ln\sum_{k=1}^M p_{\alpha,k}^2.
\end{equation}
Hence, from the concavity of the logarithm function, one obtains
\begin{equation*}
\begin{split}
\frac{1}{N}\sum_{\alpha=1}^NH(\mathcal{E}_\alpha,\rho)
&\geq-\frac{1}{N}\sum_{\alpha=1}^N\ln\sum_{k=1}^M p_{\alpha,k}^2\\&
\geq-\ln\left[\frac{1}{N}\sum_{\alpha=1}^N\sum_{k=1}^M p_{\alpha,k}^2\right]\\&
=\ln\frac{N}{C},
\end{split}
\end{equation*}
where $C$ is the index of coincidence.

\section{Separability criterion}\label{AppE}

Proposition 2 can be proven analogically to the ESIC criterion in ref. \cite{ESIC}. For a product state $\rho=\rho_A\otimes\rho_B$, the elements of the $\mathcal{P}$ matrix are simply
\begin{equation}\label{PP}
\mathcal{P}_{\alpha,k;\beta,\ell}=\Tr(\rho_AE_{\alpha,k}^A)\Tr(\rho_BE_{\beta,\ell}^B)
=p_{\alpha,k}^Ap_{\beta,\ell}^B.
\end{equation}
The trace norm reads
\begin{equation}
\begin{split}
\|\mathcal{P}\|_{\rm tr}=\Tr\sqrt{\mathcal{P}^2}&
=\sqrt{\sum_{\alpha,\beta=1}^N\sum_{k,\ell=1}^M\mathcal{P}_{\alpha,k;\beta,\ell}}\\&
=\sqrt{C_AC_B}\leq \sqrt{\widetilde{C}_A\widetilde{C}_B},
\end{split}
\end{equation}
with $\widetilde{C}_{A/B}$ being the upper bound for the index of coincidence associated with $\rho_{A/B}$. From the convexity of the trace norm, this condition follows for an arbitrary separable state.

\section{Separability criterion 2}\label{AppF}

Again, we prove the condition for a product state $\rho=\rho_A\otimes\rho_B$ and extend the end result to all separable states using linearity of trace. Starting from eq. (\ref{PP}), we write
\begin{equation}
\begin{split}
\Tr\mathcal{P}&=\sum_{\alpha=1}^N\sum_{k=1}^M\mathcal{P}_{\alpha,k;\alpha,k}
=\sum_{\alpha=1}^N\sum_{k=1}^Mp_{\alpha,k}^Ap_{\alpha,k}^B\\&\leq
\frac 12 \sum_{\alpha=1}^N\sum_{k=1}^M\Big[(p_{\alpha,k}^A)^2+(p_{\alpha,k}^B)^2\Big]
=\frac 12 (C_A+C_B)\leq\widetilde{C}.
\end{split}
\end{equation}
In the above derivation, we used the inequality of arithmetic and geometric means
\begin{equation}
\frac 1N \sum_{k=1}^Nx_k\geq \left(\prod_{k=1}^Nx_k\right)^{1/N}.
\end{equation}

\end{document}